\documentstyle[aps,preprint,epsf,epsfig,floats]{revtex}

\begin{document}

\tighten

\preprint{\font\fortssbx=cmssbx10 scaled \magstep2
\hbox to \hsize{
\hbox{\fortssbx University of Wisconsin - Madison}
\hfill$\vcenter{\hbox{\bf MADPH-96-942}
                \hbox{\bf hep-ph/9605295}
                \hbox{May 1996}}$ }
}

\title{\vspace*{.5in}
  Quantitative Tests of Color Evaporation:\\ Charmonium Production}

\author{
  J.\ F.\ Amundson\thanks{Email: amundson@phenom.physics.wisc.edu}, 
  O.\ J.\ P.\ \'Eboli\thanks{Permanent address: Instituto de 
    F\'{\i}sica, Universidade de S\~ao Paulo,
    C.P.\ 66318, CEP 05389-970 S\~ao Paulo, Brazil.
    E-mail: eboli@phenom.physics.wisc.edu}, 
  E.\ M.\ Gregores\thanks{Email: gregores@phenos.physics.wisc.edu}, 
  F.\ Halzen\thanks{Email: halzen@phenxh.physics.wisc.edu}}

\address{Department of Physics, University of Wisconsin, Madison, WI
  53706}

\maketitle

\begin{abstract}
  The color evaporation model simply states that charmonium production
  is described by the same dynamics as $D \bar D$ production, {\em
  i.e.}, by the formation of a colored $c \bar c$ pair.  Its color
  happens to be bleached by soft final-state interactions.  We show
  that the model gives a complete picture of charmonium production
  including low-energy production by proton, photon and antiproton
  beams, and high-energy production at the Tevatron and HERA. Our
  analysis includes the first next-to-leading-order calculation in the
  color evaporation model.
\end{abstract}

\newpage

\section{Introduction}

In a recent paper \cite{amundson} we pointed out that an unorthodox
prescription for the production of rapidity gaps in deep inelastic
scattering, proposed by Buchm\"uller and Hebecker \cite{bh}, suggests a
description of the production of heavy quark bound states which is in
agreement with the data. The prescription represents a reincarnation
of the ``duality'' or ``color evaporation'' model. It is very
important to study the validity of this approach as it questions the
conventional treatment of the color quantum number in perturbative
QCD.

The conventional treatment of color, {\em i.e.}, the color singlet
model, has run into serious problems describing the data on the
production of charmonium \cite{review}. While attempts to rectify the
situation exist \cite{bbl}, the color singlet model has remained the
standard by which other approaches are measured. Here we show that the
color evaporation approach, which actually predates the color singlet
approach, describes the available data well. After a discussion of the
model itself, we show how the ratio of production by antiprotons to
production by protons forms a quantitative test of the model. The
model passes the test extraordinarily well. We then show how further
quantitative tests can be performed using the the data on
photoproduction and hadroproduction total cross sections. We finally
show that the model successfully describes $p_T$ distributions at the
Tevatron.

\section{The Model}

\begin{figure}
\begin{center}
        \begin{tabular}{ccc}
                \epsfxsize=0.3\hsize
                \mbox{\epsffile{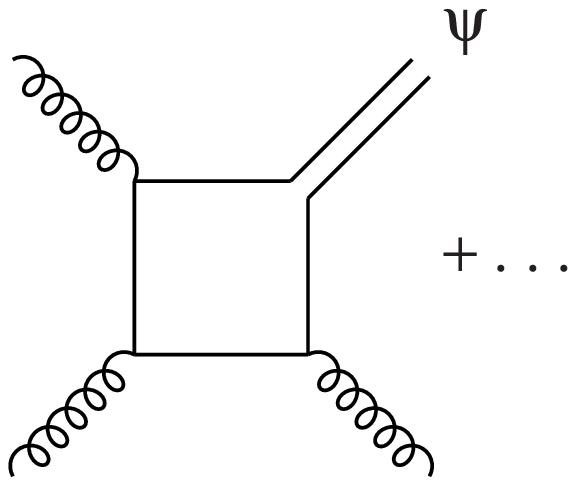}}
                &~&
                \epsfxsize=0.45\hsize
                \mbox{\epsffile{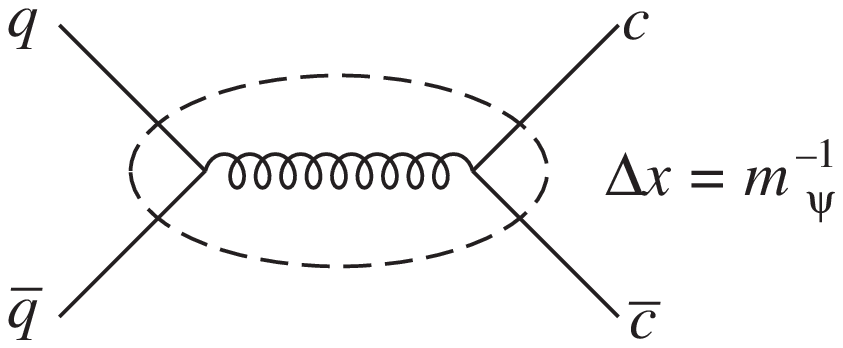}}
                \\
                (a) 
                &~&
                (b) 
                \\
        \end{tabular}
\end{center}
\caption{Typical diagrams for (a) color singlet $\psi$ production and (b) 
color evaporation $\psi$ production.}
\label{fig:csm}
\label{fig:cem}
\end{figure}

Color evaporation represents a fundamental departure from the way
color singlet states are treated in perturbation theory.  In fact,
color is ``ignored''.  Rather than explicitly imposing that the system
is in a color singlet state in the short-distance perturbative
diagrams, the appearance of color singlet asymptotic states depends
solely on the outcome of large-distance fluctuations of quarks and
gluons.  These large-distance fluctuations are probably complex enough
for the occupation of different color states to respect statistical
counting.  In other words, color is a nonperturbative phenomenon.  In
Fig.~\ref{fig:csm} we show typical diagrams for the production of
$\psi$-particles using the competing treatments of the color quantum
number.  In the diagram of Fig.~\ref{fig:csm}a, the color singlet
approach, the $\psi$ is produced in gluon-gluon interactions in
association with a final state gluon which is required by color
conservation.  This diagram is related by crossing to the hadronic
decay $\psi \rightarrow 3$ gluons.

In the color evaporation approach, the color singlet property of the 
$\psi$ is initially ignored.  For instance, the $\psi$ can be produced 
to leading order by $q\bar q$-annihilation into $c\bar c$, which is 
the color-equivalent of the Drell-Yan process.  This diagram is 
calculated perturbatively; its dynamics are dictated by short-distance 
interactions of range $\Delta x \simeq m_{\psi}^{-1}$.  It does not 
seem logical to enforce the color singlet property of the $\psi$ at 
short distances, given that there is an infinite time for soft gluons 
to readjust the color of the $c \bar c$ pair before it appears as an 
asymptotic $\psi$ or, alternatively, $D \bar D$ state.  It is indeed 
hard to imagine that a color singlet state formed at a range 
$m_{\psi}^{-1}$, automatically survives to form a $\psi$.  This 
formalism was, in fact, proposed almost twenty years ago 
\cite{cem,fh:1a,fh:1b,gor} and subsequently abandoned for no 
good reason.  

The color evaporation approach to color leads to a similar description 
of bound and open charm production.  In the color evaporation model
\begin{equation}
\sigma_{\rm onium} = \frac{1}{9} \int_{2 m_c}^{2 m_D} dm~
\frac{d \sigma_{c \bar{c}}}{dm} \; ,
\label{sig:on}
\end{equation}
and
\begin{eqnarray}
\sigma_{\rm open} &=& \frac{8}{9}  \int_{2 m_c}^{2 m_D} dm~
\frac{d \sigma_{c \bar{c}}}{dm}
+ \int_{2 m_D} dm~\frac{d \sigma_{c \bar{c}}}{dm}
\label{sig:op}
\end{eqnarray}
where the cross section for producing heavy quarks, $\sigma_{c \bar 
c}$, is computed perturbatively.  Diagrams are included order-by-order 
irrespective of the color of the $c \bar c$ pair.  The coefficients 
$\frac{1}{9}$ and $\frac{8}{9}$ represent the statistical 
probabilities that the $3\times\bar3$ charm pair is asymptotically in 
a singlet or octet state.

The model also predicts that the sum of the cross sections of all
onium states is given by Eq.~(\ref{sig:on}).
This relation is, unfortunately, difficult to test experimentally, 
since it requires measuring cross sections for {\em all} of the 
bound states at a given energy.

Other approaches similar in spirit can be found in Refs.~\cite{bbl} 
and \cite{hoyer}.  The color evaporation approach differs from 
Ref.~\cite{bbl}, the formalism of Bodwin, Braaten and Lepage (BBL), in 
the way that the $c \bar c$ pair exchanges color with the underlying 
event.  In the BBL formalism, multiple gluon interactions with the 
$c\bar c$ pair are suppressed by powers of $v$, the relative velocity 
of the heavy quarks within the $\psi$.  The color evaporation model 
assumes that these low-energy interactions can take place through 
multiple soft-gluon interactions; this implies a statistical treatment 
of color.

The color evaporation model assumes a factorization of the production 
of the $c\bar{c}$ pair, which is perturbative and process dependent, 
and the materialization of this pair into a charmonium state by a 
mechanism that is nonperturbative and process independent.  This 
assumption is reasonable since the characteristic time scales of the 
two processes are very different: the time scale for the production of 
the pair is the inverse of the heavy quark mass, while the formation 
of the bound state is associated to the time scale $1/\Lambda_{\rm 
QCD}$.  Therefore, comparison with the $\psi$ data requires knowledge 
of the fraction $\rho_\psi$ of produced onium states that materialize 
as $\psi$'s, {\em i.e.,}
\begin{equation}
\sigma_\psi = \rho_\psi \sigma_{\rm onium} \; ,
\label{frac}
\end{equation}
where $\rho_\psi$ is assumed to be a constant.  This assumption is in 
agreement with the low energy data \cite{gksssv,schuler}.  We 
demonstrated in Ref.~\cite{amundson} that simple statistical counting 
estimates of $\rho_{\psi}$, $\rho_{\chi}$, etc., accommodate all 
charmonia data to better than a factor of 2.

\section{The Tests}

We discussed several qualitative tests of the color evaporation 
picture in Ref.~\cite{amundson}.  One such test that we did not 
mention is the polarization of produced charmonium.  In the framework 
of the color evaporation model the multiple soft gluon exchange 
destroys the initial polarization of the heavy quark pair 
\cite{mirkes}.  This fact is in agreement with the measurements of the 
$\psi$ polarization made in fixed-target $\pi^-N$ \cite{pin} and 
$\bar{p}N$ \cite{pbarn} reactions.  The color singlet model 
fails to describe this feature of charmonium production \cite{tang} 
since it predicts that $\psi$'s are produced transversely polarized.  
The predictions in the literature for polarization in the BBL 
formalism have been somewhat controversial\cite{braatenchen}.

One of the most striking features of color evaporation is that the 
production of charmonium at low energies is dominated by the 
conversion of a colored gluon into a $\psi$, as in 
Fig.~\ref{fig:cem}b.  In the conventional treatment, where color 
singlet states are formed at the perturbative level, 3 gluons (or 2 
gluons and a photon) are required to produce a $\psi$.  The result is 
that in the color evaporation model $\psi$'s are hadroproduced not 
only by gluon-gluon initial states, but also via quark-antiquark 
fusion.  In the color singlet approach such diagrams only appear at 
higher orders of perturbation theory and their contribution is small.

\begin{figure}
\begin{center}
\mbox{\epsfig{file=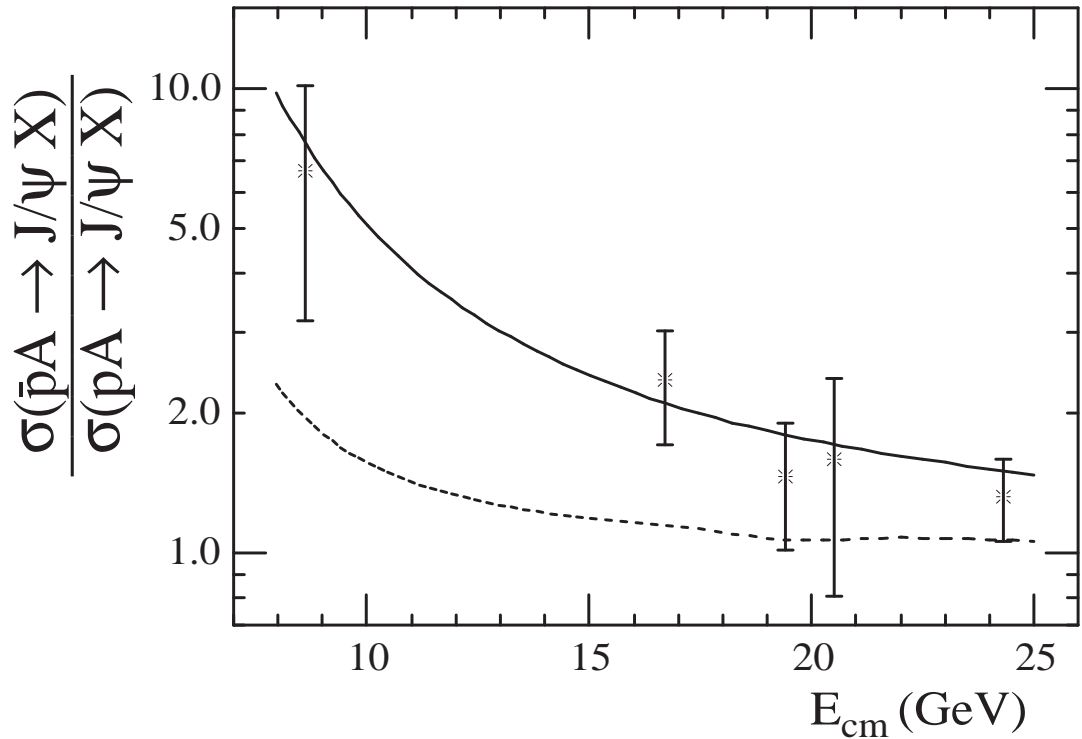,width=.6\linewidth,angle=-90}}
\end{center}
\caption{Ratio of the cross sections for the production of $J/\psi$ 
by proton and antiproton beams in the color evaporation model (solid 
line) and the color singlet model (dashed line) as a function of the
center-of-mass energy.  Data taken from 
Ref.~\protect\cite{exp:ratio}.}
\label{fig:part-anti}
\end{figure}

We can distinguish between the two pictures experimentally by 
comparing the production of charmonium by proton beams with that from 
antiproton beams.  The color evaporation model predicts an enhanced 
$\psi$ cross section for antiproton beams, while the color singlet 
model predicts the same cross section for the production of $\psi$'s, 
whether we use proton or antiproton beams.  The prediction of an 
enhanced $\bar p$ yield compared to $p$ yield at low energies is 
obviously correct: the ratio of antiproton and proton production of 
$\psi$'s exceeds a factor 5 close to threshold.  (See 
Fig.~\ref{fig:part-anti}.)  In fact, it has been known for some time 
that $\psi$'s are predominantly produced by $q\bar q$ states 
\cite{fh:1a,fh:1b,gor}.  Nonetheless, we should note that for 
sufficiently high energies, gluon initial states will eventually 
dominate because they represent the bulk of soft partons.  This can be 
seen in Fig.~\ref{fig:part-anti} where the ratio gets close to unity for 
center-of-mass energies as low as 25 GeV.

\begin{figure}
\begin{center}
\mbox{\epsfig{file=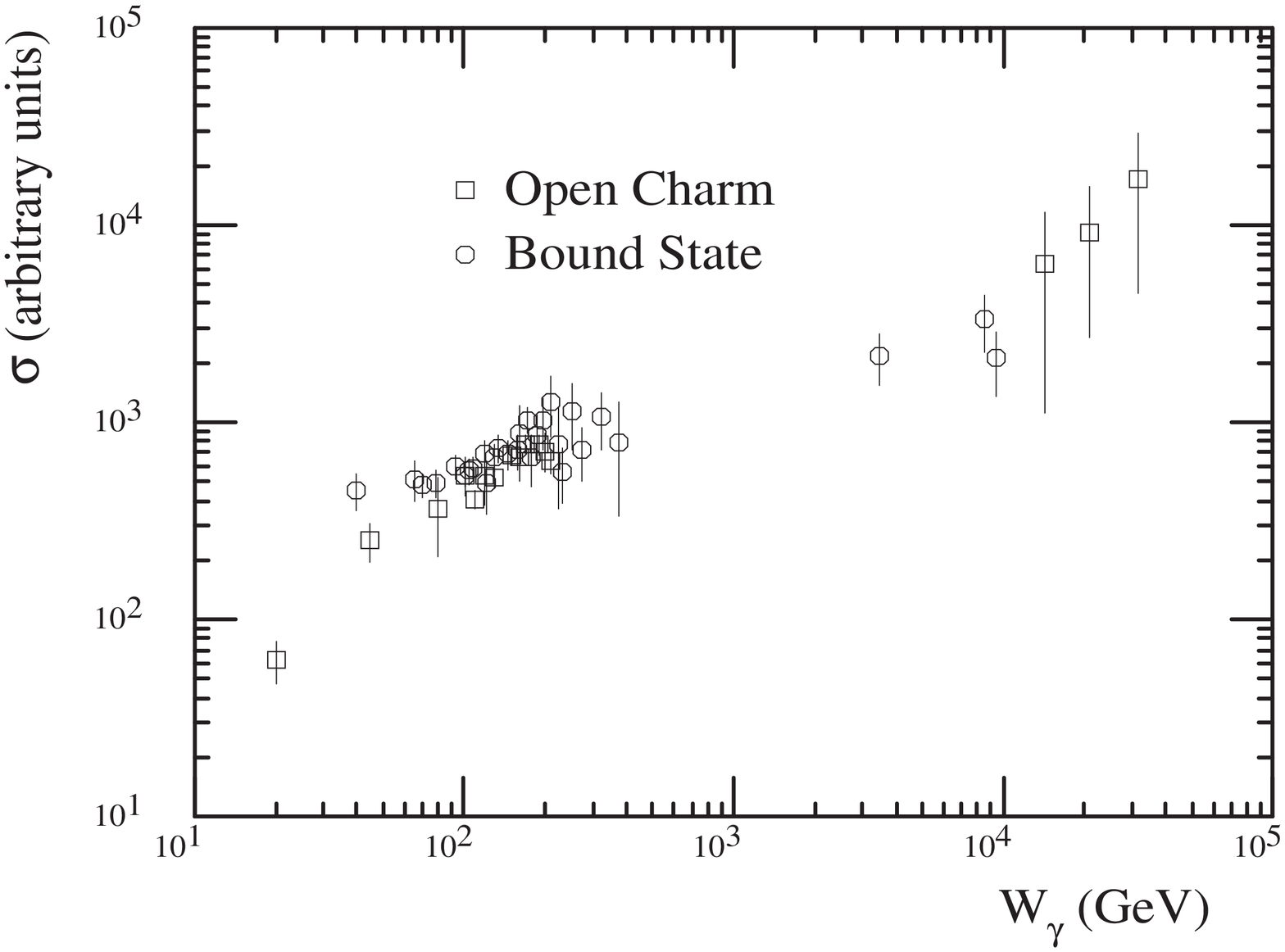,angle=-90,width=.8\linewidth}}

\end{center}
\caption{Photoproduction data \protect\cite{data1,data2} as a function
  of the photon energy in the hadron rest frame, $W_\gamma$.  The
  normalization has been adjusted to show the similar shapes of the
  data.}
\label{fig:justdata}
\end{figure}

\begin{figure}
\begin{center}
\mbox{\epsfig{file=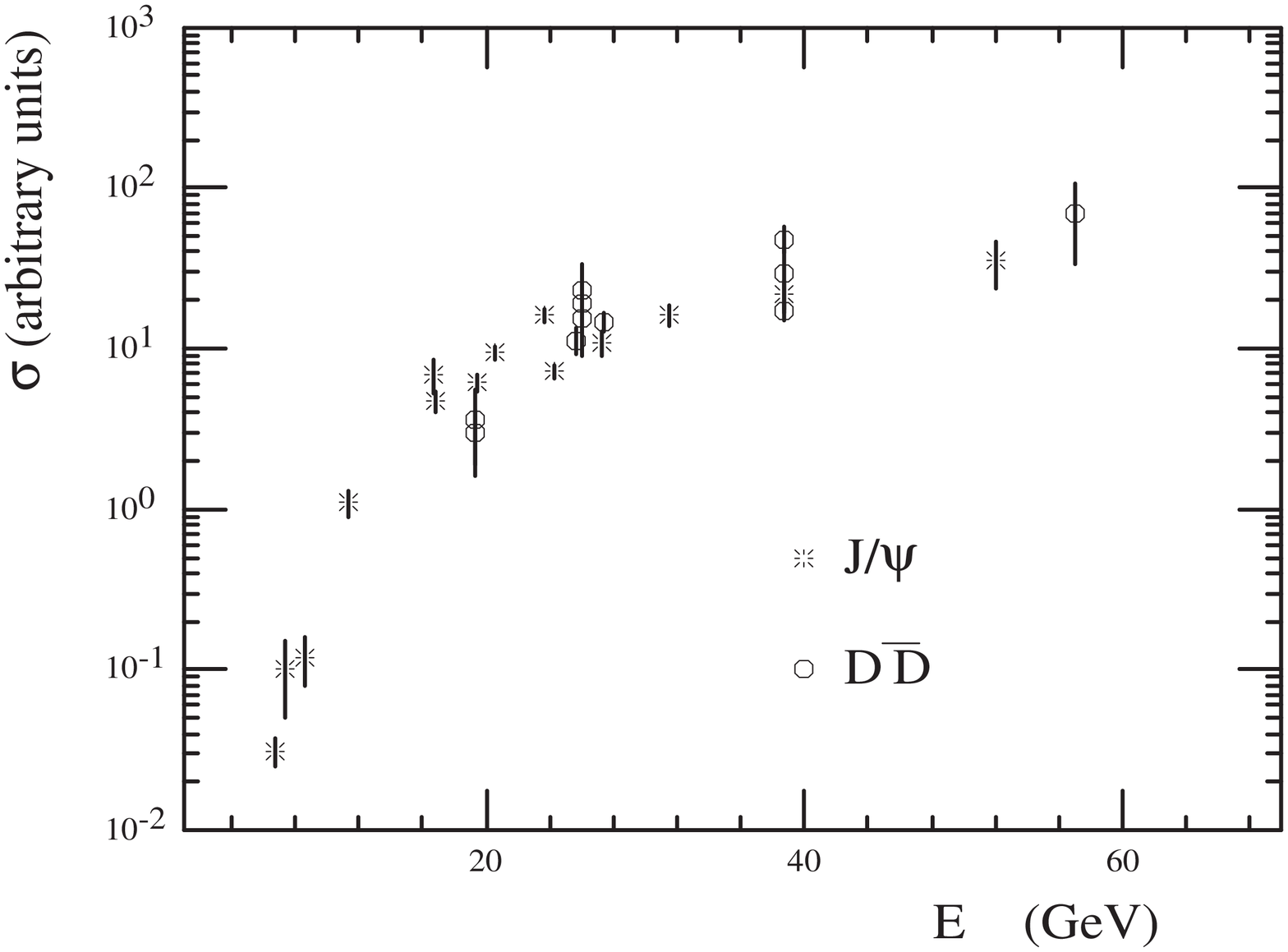,width=.8\linewidth,angle=-90}}
\end{center}
\caption{Hadroproduction data as a function of the center-of-mass 
energy, $E_{cm}$.  Data taken from 
Refs.~\protect\cite{exp:psi,exp:charm}.  Once again, the normalization 
has been adjusted to show the similar shapes of the data.}
\label{fig:hadro-data}
\end{figure}

The merits of the color evaporation approach can be first appreciated 
by studying the data in calculation-independent ways.  The 
factorization of the production of the $c\bar{c}$ pairs and the 
formation of onium states implies that the energy dependence and 
kinematic distributions of the measured cross section for the 
different bound states should be equivalent.  Moreover, in the 
approximation $m_{c}\approx m_{D}$ this equivalence extends to the 
production of open $D \bar D$ pairs.  In Fig.~\ref{fig:justdata} we 
display charm photoproduction data for both open charm and bound state 
production with common normalization in order to show their identical 
energy behavior. 
In Fig.~\ref{fig:hadro-data} we display charm 
hadroproduction data in a similar fashion.

Further quantitative tests of color evaporation are made possible by 
the fact that the factor $\rho_\psi$ is the same in hadro- and 
photoproduction.  Once $\rho_\psi$ has been empirically determined for 
one initial state, the cross section is predicted without free 
parameters for the other.  We show next that color evaporation passes 
this test, quantitatively accommodating all measurements, including 
the high energy Tevatron and HERA data, which have represented a 
considerable challenge for the color singlet model.

\begin{figure}
\begin{center}
\mbox{\epsfig{file=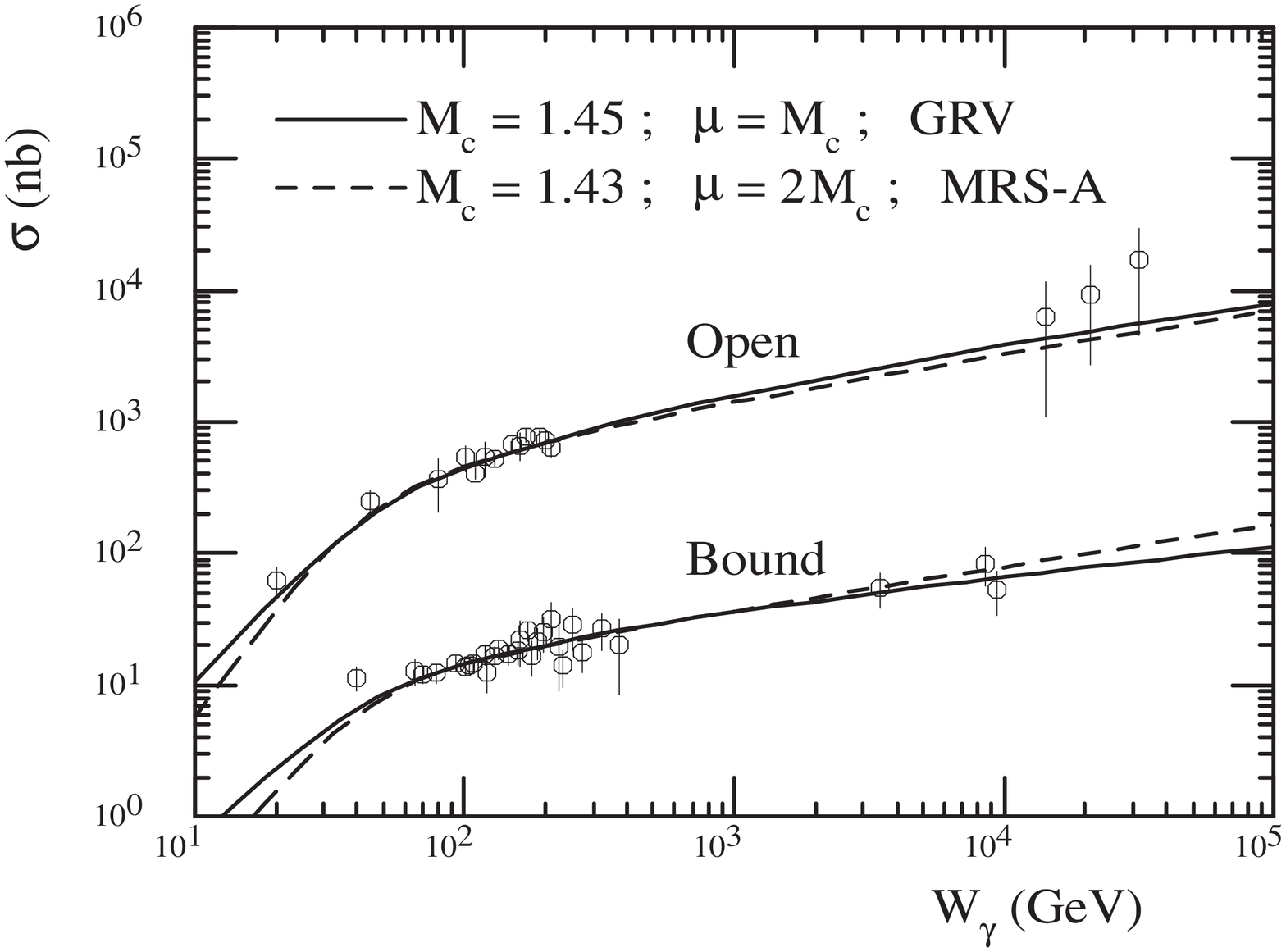,width=.8\linewidth,angle=-90}}
\end{center}
\caption{
  Photoproduction data \protect\cite{data1,data2} and the predictions
  of the color evaporation model at next-to-leading order as a
  function of the photon energy in the hadron rest frame, $W_\gamma$.
  The normalizations in this figure are absolute.}
\label{fig:photopro}
\end{figure}

In Fig.~\ref{fig:photopro} we compare the photoproduction data with
theory, using the NLO perturbative QCD calculation of charm pair
production from Ref.~\cite{nlo}.  The solid line is the NLO prediction
for open charm production using the GRV HO distribution function and
the scale $\mu = m_c = 1.45$~GeV. The dashed line is the prediction
using the MRS~A distribution function with the scale $\mu = 2 m_c =
2.86$~GeV. We obtained these charm quark masses from the best fit to
the data for each structure function.  For the $J/\psi$ production
data we employed the parameters used for describing open charm, and
determined the fragmentation factor $\rho_\psi$ to be 0.50 using GRV
HO, or 0.43 using MRS~A. Note that the factor $\rho_\psi$ possesses a
substantial theoretical uncertainty due to the choice of scales and
parton distribution functions.  We conclude the photoproduction of
$J/\psi$ and $D\bar{D}$ is well described by the color evaporation
model.  This reaction determines the only free parameter,
$\rho_\psi\approx 0.5$.

\begin{figure}
\begin{center}
\mbox{\epsfig{file=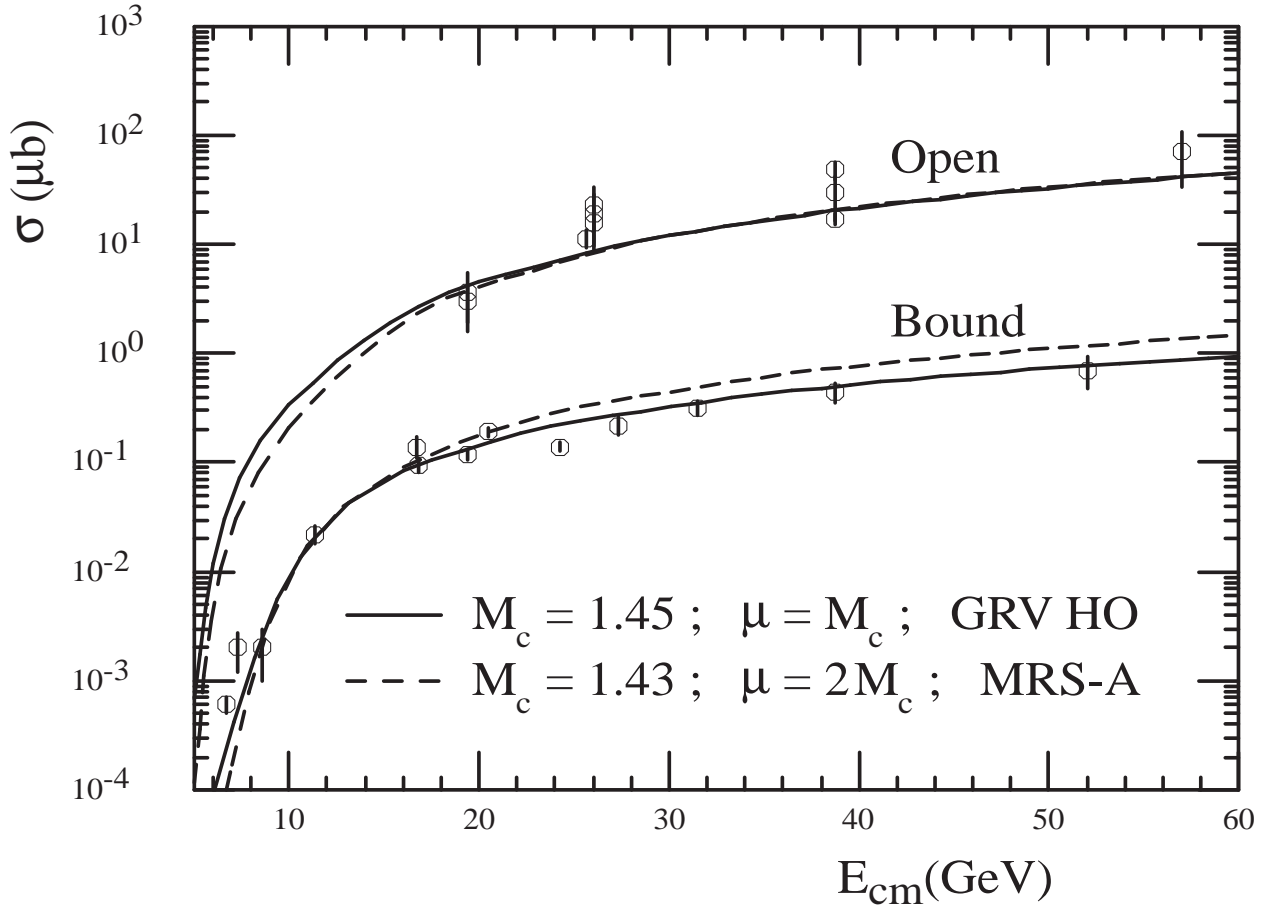,width=.8\linewidth,angle=-90}}
\end{center}
\caption{
  Hadroproduction data \protect\cite{exp:psi,exp:charm} and the
  predictions of the color evaporation model at next-to-leading order
  as a function of the center-of-mass energy, $E_{cm}$.  The curve
  for bound state production is an absolutely normalized,
  parameter-free prediction of the color evaporation model.}
\label{fig:hadro-fit}
\end{figure}

At this point the predictions of the color evaporation model for 
hadroproduction of $\psi$ are completely determined, up to higher 
order QCD corrections.  These can be determined by fitting the 
hadroproduction cross section of $D \bar{D}$ pairs with a global $K$ 
factor.  This factor is subsequently used to correct the $\psi$ 
prediction.  In Fig.~\ref{fig:hadro-fit} we compare the color 
evaporation model predictions with the data.  In order to fit the 
$D\bar{D}$ cross section with the NLO result for the production of 
$c\bar{c}$ pairs we introduced a factor $K=1.27$ (1.71) for the GRV HO 
(MRS~A) distribution function with the same scale $\mu$ and charm 
quark mass determined by photoproduction.  Inserting this $K$ factor 
and the constant $\rho_\psi$ obtained from photoproduction into 
Eq.~(\ref{sig:on}), we predict $\psi$ hadroproduction without any free 
parameters.  We conclude from Fig.~\ref{fig:hadro-fit} that the color 
evaporation model describes the hadroproduction of $\psi$ very 
accurately.

\begin{figure}
\begin{center}
\mbox{\epsfig{file=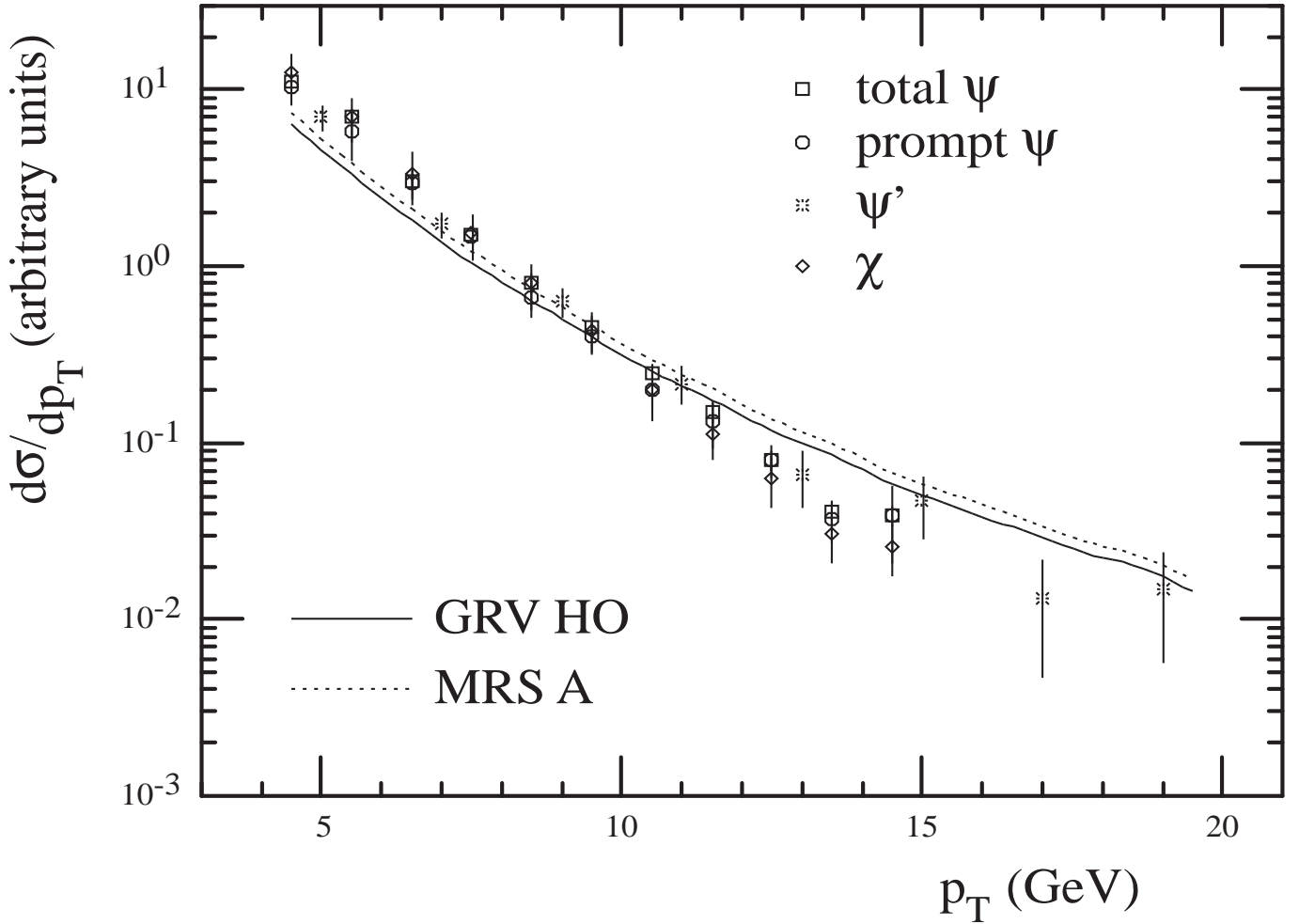,width=.8\linewidth,angle=-90}}
\end{center}
\caption{Data from the CDF Collaboration \protect\cite{exp:pt}, shown 
with arbitrary normalization.  The curves are the predictions of the 
color evaporation model at tree level, also shown with arbitrary 
normalization. The normalization is correctly predicted within a 
K-factor of 2.2.}
\label{fig:cdf-pt}
\end{figure}

As a final test, we turn to $p_{T}$ distributions at the Tevatron.
The CDF collaboration has accumulated large samples of data on the
production of prompt $\psi$, $\chi_{c{\tiny J}}$, and $\psi^\prime$
\cite{exp:pt}.  This data set allows a detailed study of the $p_T$
distribution of the produced charmonium states.  Since all the
charmonium states share the same production dynamics in the color
evaporation model, their $p_T$ distributions should be the same, up to
a multiplicative constant.  This prediction is confirmed by the CDF
data, as we can see in Fig.~\ref{fig:cdf-pt}.

In order to obtain the theoretical prediction, we computed the
processes $g + g \to [c\bar{c}] + g$, $q + \bar{q} \to [c\bar{c}] +
g$, and $g + q \to [c\bar{c}] + q$ at tree level using the package
MADGRAPH \cite{tim}.  This calculation should give a reliable estimate
of the $p_T$ distribution at large values of $p_T$.  We require that
the $c\bar{c}$ pair satisfy the invariant mass constraint on
Eq.~(\ref{sig:on}).  Our results are shown in Fig.~\ref{fig:cdf-pt},
which required a $K$ factor of $2.2$ in order to fit the total $\psi$
production.  As we can see from this figure, the color evaporation
model describes the general features of the $p_T$ distribution of the
different charmonium states.  Furthermore, we should keep in mind that
the factor $K$ is, in general, $p_T$ dependent and that it is usually
larger at low $p_T$ \cite{vogt}.  Higher order corrections such as
soft-gluon resummation are expected to tilt our lowest order
prediction, bringing it to a closer agreement with the data
\cite{sean}.

\section{Conclusions}

The color evaporation model died an untimely death. In its current
reincarnation its qualitative validity can be proven directly from the
available experimental data by taking ratios of production by
different beams and production of different particles. Moreover, we
have shown that its validity extends to the quantitative regime once
we use next-to-leading-order QCD calculations for the photo- and
hadroproduction of $\psi$ and $D\bar{D}$ pairs.

The color evaporation model explains all available data on $p_T$
distribution and energy-dependence of the cross section for the
production of $\psi$ in all the available energy range. Moreover, it
sheds some light in the relation between the production of charmonium
states and $D\bar{D}$ pairs. 

\acknowledgments 

We would like to thank P.~Nason and R.~K.~Ellis for providing us the
code containing the next-to-leading-order QCD cross sections.  We
would also like to thank S.~Fleming for his insight.  This research
was supported in part by the University of Wisconsin Research
Committee with funds granted by the Wisconsin Alumni Research
Foundation, by the U.S.\ Department of Energy under grant
DE-FG02-95ER40896, and by Conselho Nacional de Desenvolvimento
Cient\'{\i}fico e Tecnol\'ogico (CNPq).

\end{document}